\documentclass[fleqn,10pt]{wlscirep}
\title{Precision imaging of 4.4 MeV gamma rays using a 3-D position sensitive Compton camera}

\author[1]{Ayako Koide}
\author[1,*]{Jun Kataoka}
\author[1]{Takamitsu Masuda}
\author[1]{Saku Mochizuki}
\author[1]{Takanori Taya}
\author[1]{Koki Sueoka}
\author[1]{Leo Tagawa}
\author[1]{Kazuya Fujieda}
\author[1]{Takuya Maruhashi}
\author[1]{Takuya Kurihara}
\author[2]{Taku Inaniwa}
\affil[1]{Waseda University, Graduate School of Advanced Science and Engineering, Tokyo, Japan}
\affil[2]{National Institute of  Radiological Sciences, QST, Department of Accelerator and Medical Physics, Chiba, Japan}

\affil[*]{kataoka.jun@waseda.jp}

\begin{abstract}
Imaging of nuclear gamma-ray lines in the 1-10~MeV range is far from being established in both medical and physical applications.  In proton therapy, 4.4 MeV gamma rays are emitted from the excited nucleus of either $^{12}$C$^{*}$ or $^{11}$B$^{*}$ and are considered good indicators of dose delivery and/or range verification. Further, in gamma-ray astronomy, 4.4~MeV gamma rays are produced by cosmic ray interactions in the interstellar medium, and can thus be used to probe nucleothynthesis in the universe. In this paper, we present a high-precision image of 4.4~MeV gamma rays taken by newly developed 3-D position sensitive Compton camera (3D-PSCC). To mimic the situation in proton therapy, we first irradiated water, PMMA and Ca(OH)$_2$ with a 70~MeV proton beam, then we identified various nuclear lines with the HPGe detector. The 4.4~MeV gamma rays constitute a broad peak, including single and double escape peaks. Thus, by setting an energy window of 3D-PSCC from 3 to 5~MeV, we show that a gamma ray image sharply concentrates near the Bragg peak, as expected from the minimum energy threshold and sharp peak profile in the cross section of $^{12}$C(p,p)$^{12}$C$^{*}$.
\end{abstract}
\begin{document}

\flushbottom
\maketitle
%
%
\thispagestyle{empty}


\section*{Introduction}
Gamma ray imaging techniques are widely used in various fields of nuclear medicine, industries, nuclear and elementary particle physics, and also in high energy astrophysics. For example, single photon emission computed tomography (SPECT) is a tomographic gamma ray imaging technique in which a flow tracer is tagged with a radionuclide. A popular tracer is $^{99{\rm m}}$Tc, which emits 140~keV photons. But other tracers, such as $^{111{\rm m}}$In (171~keV, 245~keV) and $^{57}$Co (122~keV), are also used to visualize concentrations in different organs\cite{Khalil:2011dg}. The SPECT imaging is, however, limited to use with photon energies below 300~keV because high resolution images cannot be obtained as higher energy gamma rays penetrate the collimator\cite{Rosenthal:1995dg}. Positron emission tomography (PET) allows visualization of FDG (fluoro-deoxy-glucose) tagged with positron emitters, most often $^{18}$F\cite{Kubota:2005dg}. Owing to the use of pair annihilation of 511~keV gamma rays, a PET scanner does not need a collimator and is widely used to find tumors and diagnose Alzheimer' disease\cite{Coleman:2005dg}. In both imaging techniques, photo-absorption is the dominant process within a detector, thus a dense high speed scintillator is generally used. In contrast, very high energy gamma rays (greater than 10~MeV) are often an active research target in modern astronomy and elementary particle physics. At such energies, the dominant interaction between photons in the material is pair production, therefore particle tracking systems like silicon strip detectors (SSD\cite{Weilhammer:2000dg}) or scintillation fibers are commonly used to determine the incident direction of gamma rays\cite{Torii:2015dg,Knodlseder:2016dg}. For example, large area telescopes (LAT) in the Fermi gamma ray space telescope launched in 2008, carried $\sim$10,000 SSDs\cite{Atwood:2009dg}. Fermi-LAT successfully provided a  sky map between 100~MeV and more than 10~GeV with a typical angular resolution of $\sim$0.6$^{\circ}$ as measured at 1~GeV\cite{Acero:2015dg}.

In this context, gamma ray imaging in the 1$-$10~MeV energy range has lagged by a few decades owing to difficulties in the optimum detector configuration. In fact, gamma rays with energy greater than 1~MeV are difficult to collimate like in the SPECT scanner, whereas the energy is too low to expect a sufficient cascade of $e^{-}$$e^{+}$ for tracking detectors like the SSD. Now the dominant interaction is Compton scattering, thus the kinematics of both electrons and photons must be considered in order to the determine incident direction of photons\cite{shon:1973dg}. Despite such difficulties, however, imaging of 1$-$10~MeV gamma rays may provide fruitful scientific/technical outputs because most nuclear gamma ray lines are narrowly concentrated in this particular energy band. For example, in proton therapy, incident protons interact with atoms in the patient's body, and then promptly emit nuclear gamma rays, such as 4.4~MeV lines from $^{12}$C$^{*}$ and $^{11}$B$^{*}$, and the 6.2~MeV line from $^{15}$O$^{*}$\cite{Lopes:2015dg,Golnik:2016dg,Hilaire:2016dg,Kelleter:2017dg,Koide:2018dg}. These gamma rays are supposed to be a new tool for online monitoring of dose delivery and/or proton range verification\cite{Min:2006dg,Knopf:2013dg,Kurosawa:2012dg}. The online monitoring technique may replace the current offline monitor based on PET scanners, or even online PET scanners that visualize 511~keV gamma rays\cite{Nishio:2010dg,Taya:2016dg,Taya:2017dg}. Even in high energy astrophysics, various gamma ray lines are anticipated in solar flares, planetary atmospheres including the Earth\cite{Kozol:2002dg}, radioactive elements on the lunar surface\cite{Kobayashi:2010dg}, and even through the nuclear interaction between cosmic rays and the interstellar medium. Actually, various gamma lines have been frequently observed in solar flares\cite{Vestrand:1999dg}.
Further, 1.156~MeV nuclear gamma rays from $^{44}$Ti were detected from nearby young supernova remnants of Cassiopeia A, where the half-life of $^{44}$Ti is 48.2$\pm$0.9~yrs\cite{Iyudin:1994dg}. Also, 1.809~MeV gamma rays from $^{26}$Al,  which is a radioactive isotope whose half-time is ~Myr, was already detected from the Galactic plane over 30~years ago, showing direct evidence that nucleosynthesis is ongoing in the star-forming region in our Galaxy\cite{Diehl:1995dg}. Similarly, various emission lines, including 4.4~MeV and 6.2~MeV lines, are expected from the Galactic center region\cite{Dogiel:2009dg}, but none of them have been detected so far.

A Compton camera utilizes Compton kinematics to visualize gamma rays ranging from sub-MeV to more than 10~MeV\cite{shon:1973dg}. Unlike a traditional pinhole camera\cite{Beekman:2007dg} or an imager using a coded mask aperture\cite{Caroli:1987dg}, a Compton camera does not need a heavy collimator or shield. In general, this achieves high sensitivity and wide field of view. Although variety exists in the choice of detector materials\cite{shon:1993dg,Tanimori:2004dg,Takahashi:2004dg,Kataoka:2013dg,Motomura:2013dg}, a Compton camera generally consists of a scatterer, wherein the incident gamma rays first deposit a fraction of energy, and an absorber in which the remaining energy carried by scattered photons is absorbed. Obviously, the energy of incident gamma rays are the sum of those deposited in the scatterer and absorber. The arrival direction can be determined (or at least, constrained) by measuring both the scattering and absorbing positions and energies within the detector\cite{Tanimori:2017dg}. The concept of a Compton camera was first proposed over 40~years ago and targeted both the medical and astrophysical applications\cite{shon:1973dg,Tod:1974dg}, but vast progress has been made only recently. This was triggered by a nuclear accident in the Fukushima Daiichi Nuclear Plant in 2011. While the radioactive material distributed in the accident was mainly $^{137}$Cs, which emits 662~keV gamma rays, various attempts were made to visualize sub-MeV to MeV gamma-rays quickly and accurately with the most recent detector technologies\cite{Kataoka:2013dg,Kishimoto:2014dg,Wahl:2015dg,Takeda:2015dg,Tomono:2017dg,Mochizuki:2017dg,Andrew:2017dg}.

In this paper, we developed a novel 3-D position sensitive Compton camera (3D-PSCC) that enables gamma ray imaging of the 1$-$10~MeV range. As a first step toward future applications, we started with measurements assuming the online proton therapy monitoring. We irradiated various phantoms with a pencil beam of 70~MeV protons, and we obtained the energy spectra of prompt gamma rays below 5 MeV as measured with a high-purity Germanium (HPGe) detector. Among the various nuclear emission lines identified, we suggest that broad 4.4~MeV lines associated with $^{12}$C$^{*}$ and $^{11}$B$^{*}$ are most suitable for proton range verification in terms of minimum energy threshold and sharp nuclear cross section structure. Subsequently, the idea is supported by detailed 1-D profile simulations of various prompt gamma rays. By using multiple PMMA slab phantoms to effectively extend the proton range, we demonstrate, for the first time in experiment, that 4.4~MeV gamma rays sharply concentrate near the Bragg peak.

\begin{figure}[ht]
\centering
\includegraphics[width=0.6\linewidth]{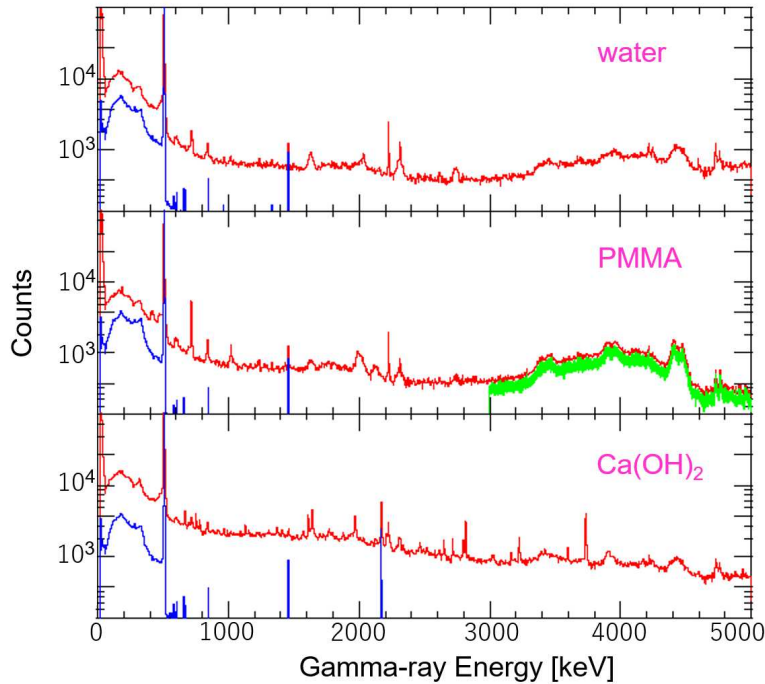}
\caption{Gamma-ray spectra for 70~MeV proton irradiation on various phantoms, obtained with an HPGe detector for on-beam ($red$) and off-beam ($blue$) conditions. ($upper$) Water, ($middle$) PMMA, and ($bottom$) Ca(OH)$_2$. The $thick$ $green$ curve in the PMMA spectra indicates an energy window applied for 4.4~MeV gamma ray imaging using the 3D-PSCC.}
\label{fig:HPGEspec}
\end{figure}

\begin{table}[ht]
\centering
\begin{tabular}{|c|c|c|c|c|}
\hline
Energy [keV] & nuclear reaction & water & PMMA & Ca(OH)$_2$\\
\hline
511  & $^{12}$C(p,x)$^{11}$C, $^{16}$O(p,x)$^{15}$O, $^{16}$O(p,x)$^{13}$N, $^{16}$O(p,x)$^{11}$C  & \checkmark & \checkmark & \checkmark\\ 
718  & $^{12}$C(p,x)$^{10}$B$^{*}$, $^{16}$O(p,x)$^{10}$B$^{*}$,$^{12}$C(p,x)$^{10}$C $\rightarrow$ $^{10}$B$^{*}$ & \checkmark & \checkmark & \checkmark\\
1022 & $^{12}$C(p,x)$^{10}$B$^{*}$, $^{12}$C(p,x)$^{10}$C $\rightarrow$ $^{10}$B$^{*}$,  $^{16}$O(p,x)$^{10}$B$^{*}$ &  & \checkmark & \\
1635 & $^{16}$O(p,x)$^{14}$N$^{*}$ & \checkmark &  & \\
2000 & $^{12}$C(p,x)$^{11}$C$^{*}$, $^{16}$O(p,x)$^{11}$C$^{*}$ &  & \checkmark & \\
2225 & neutron capture & \checkmark & \checkmark & \checkmark \\
2313 & $^{16}$O(p,x)$^{14}$N$^{*}$ & \checkmark & \checkmark & \checkmark \\
3736 & $^{40}$Ca(p,p)$^{40}$Ca$^{*}$ & &  & \checkmark \\
4438/4444 & $^{12}$C(p,p)$^{12}$C$^{*}$, $^{12}$C(p,x)$^{11}$B$^{*}$, $^{16}$O(p,x)$^{12}$C$^{*}$ & \checkmark & \checkmark & \checkmark \\ 
\hline
\end{tabular}
\caption{\label{tab:Table.1}
Nuclear gamma ray lines\cite{Kozol:2002dg} detected in the HPGe measurements and subsequently confirmed by PHITS\cite{Sato:2018dg} (ver.2.93) simulations.}
\end{table}

\begin{figure}[ht]
\centering
\includegraphics[width=0.7\linewidth]{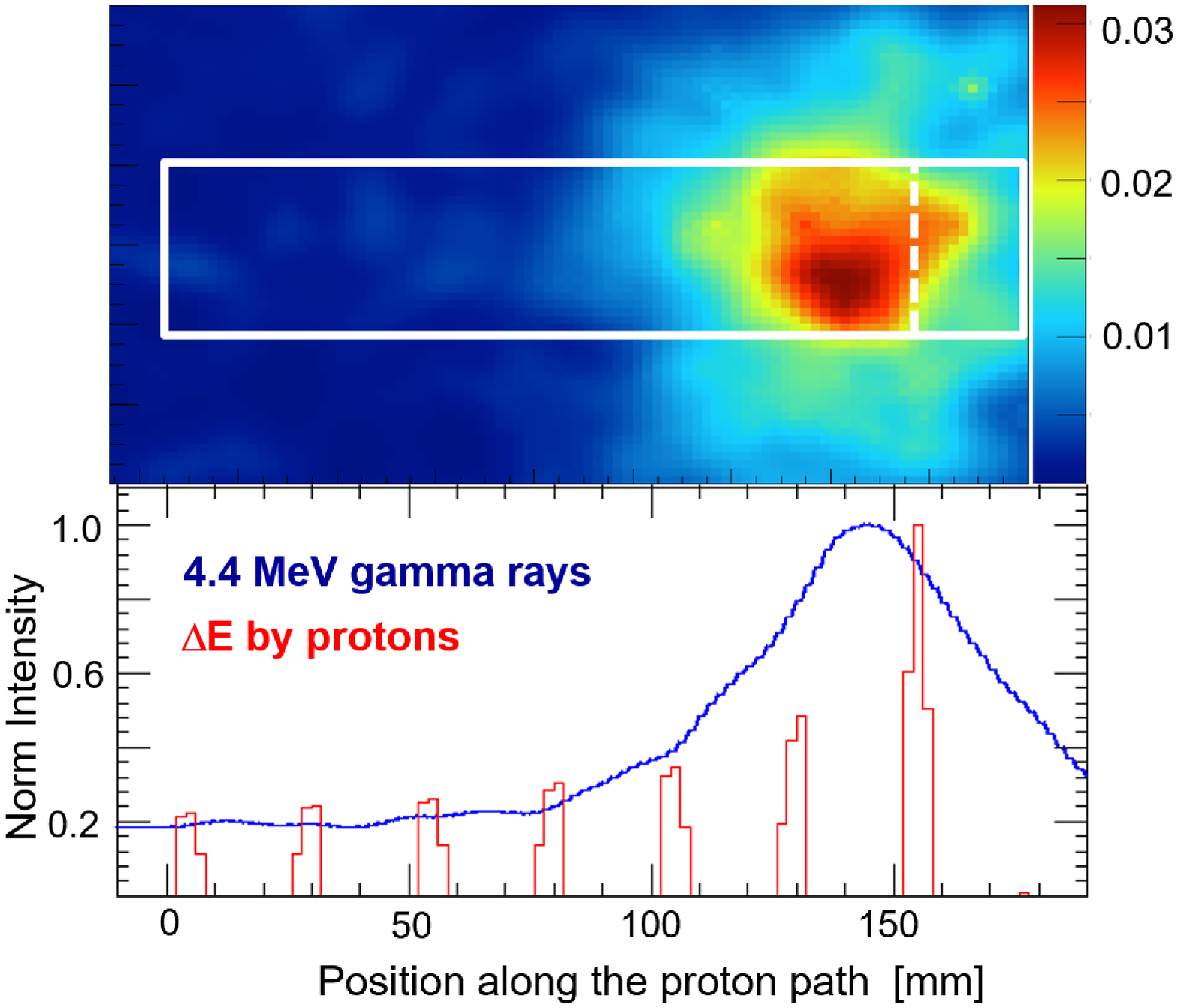}
\caption{
($upper$) The first experimental image of the 4.4~MeV prompt gamma rays reconstructed from the 3D-PSCC measurement. The $white$ $box$ indicates the region where the eight PMMA slab phantoms (0.5~cm width) were placed, and the $dotted$ $line$ marks the position of the Bragg peak for 70~MeV protons. ($lower$) 1-D projection along the proton path (shown in $blue$) of the 4.4~MeV gamma-ray image presented above, and a comparison of the energy deposited by protons ($red$ $histogram$) obtained from the PHITS simulation.
}
\label{fig:BPexp}
\end{figure}

\section*{Results}
\subsection*{Prompt Gamma-ray Spectra}
To investigate various prompt gamma rays that could be emitted from the patient's body during proton therapy, we irradiated water, PMMA, and Ca(OH)$_{2}$ phantoms with 70~MeV protons and measured the gamma ray spectra with an HPGe detector in both the on-beam and off-beam conditions. Fig.1 shows the resulting spectra for these three phantoms during proton irradiation (on-beam; $red$ $line$) as compared with the background spectra, taken soon after the beam is turned off (off-beam; $blue$ $line$). All the presented spectra are corrected for efficiency of HPGe detector, which decreases as the incident gamma ray energy increases. In all cases, various emission lines from nuclear reactions and a broad continuum due to inelastic scattering of protons are clearly visible up to a maximum energy end of 5~MeV. Table.1 summarizes the identification of most strong lines seen in the online spectra, based on a detailed comparison with PHITS simulations\cite{Sato:2018dg} (ver.2.93). Note that the very broad structures most  clearly seen in the PMMA spectra between 3~MeV and 5~MeV (shown as a $thick$ $green$ $line$), is the combination of prompt gamma rays emitted from $^{12}$C$^{*}$ (4438~keV) and $^{11}$B$^{*}$ (4444~keV) and their single or double escape peaks. In contrast, the only prominent line in the off-beam spectrum is the 511~keV annihilation line from a positron emitter like $^{15}$O. As noted above, these 511~keV gamma rays are being widely used for offline monitoring in proton therapy. 

\subsection*{Precision imaging of 4.4~MeV Gamma rays}
Among the various emission lines listed above, we targeted 4.4~MeV gamma rays for future on-line monitoring in proton therapy. Note that the 1-D profile of 4.4~MeV gamma rays along the proton beam is reported by several other groups mainly by using a slit collimator\cite{Min:2006dg,Lopes:2015dg,Kelleter:2017dg}, but no experimental results on 2-D images have been reported so far, particularly using a Compton camera. For precision measurement of a 4.4~MeV gamma ray image and its detailed comparison with the energy deposited by protons, we developed 3-D PSCC that features the 1$-$10~MeV range. The expected angular resolution is 6.4$^{\circ}$ (FWHM) as measured at 4.4~MeV. Fig.2 ($upper$) shows the 4.4~MeV gamma ray image, and Fig.2 ($bottom$) shows a comparison between the 1-D projection along the beam and the energy deposited by incident protons. Clearly, the peak position of the 4.4~MeV gamma rays image agrees well but is slightly shifted by a few mm, which is again consistent with the PHITS simulation. 

\begin{figure}[ht]
\centering
\includegraphics[width=0.9\linewidth]{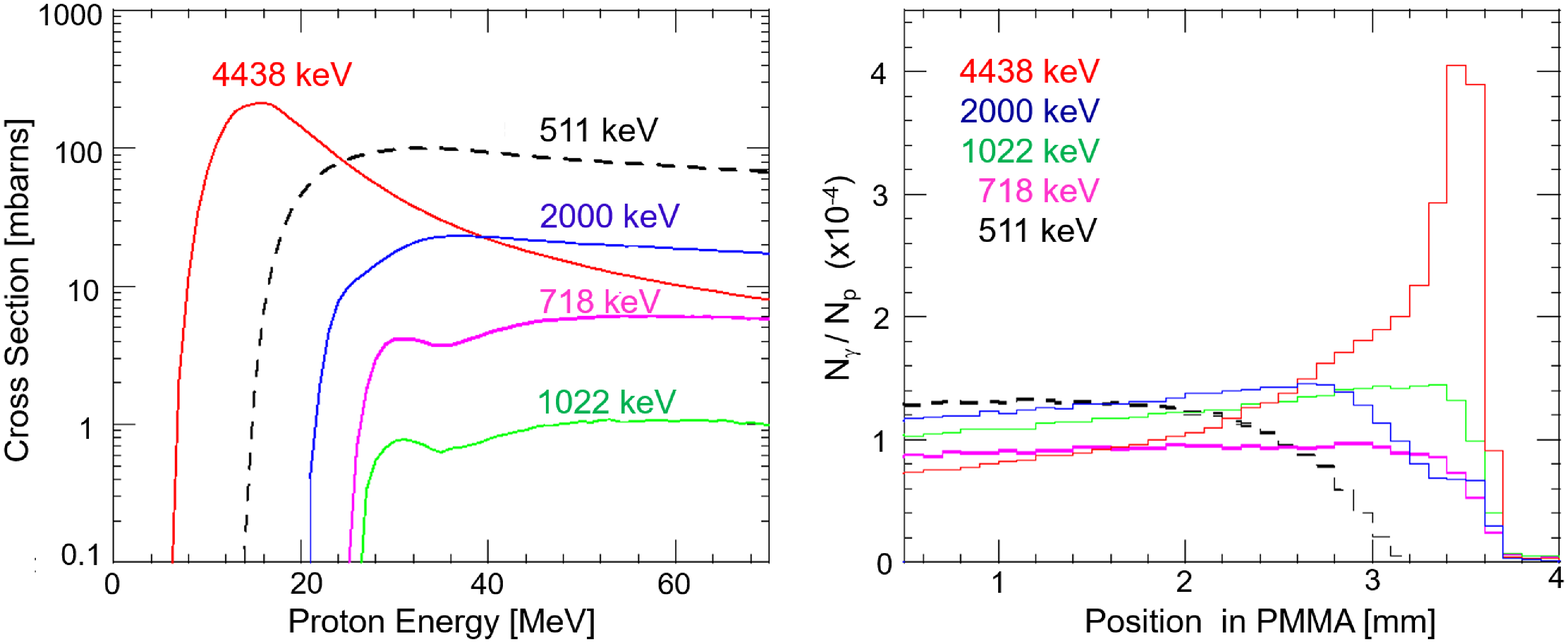}
\caption{
 ($left$) Nuclear cross sections for the major prompt gamma ray emissions as a function of proton energy. 718~keV ($magenta$; $^{12}$C(p,x)$^{10}$B$^{*}$), 1022~keV ($green$; $^{12}$C(p,x)$^{10}$B$^{*}$), 2000~keV ($blue$; $^{12}$C(p,x)$^{11}$C$^{*}$), and 4438~keV ($red$; $^{12}$C(p,p)$^{12}$C$^{*}$) as calculated from the PHITS\cite{Sato:2018dg} ver.2.93 simulation code. The primary nuclear reaction that produces positron emitters, $^{16}$O(p,x)$^{15}$O, is also plotted in $dashed$ $black$ $line$ as a reference. ($right$) Comparison of the simulated 1-D distribution of the major prompt gamma rays along the proton path in the PMMA phantom obtained with the PHITS simulation. 511~keV ($black$), 718~keV ($magenta$), 1022~keV ($green$), 2000~keV ($blue$), and 4438~keV ($red$). The incident proton energy is 70~MeV.
The vertical axis represents the number of emitted gamma rays as normalized with the incident number of protons.}
\label{fig:cross}
\end{figure}

\begin{figure}[ht]
\centering
\includegraphics[width=0.8\linewidth]{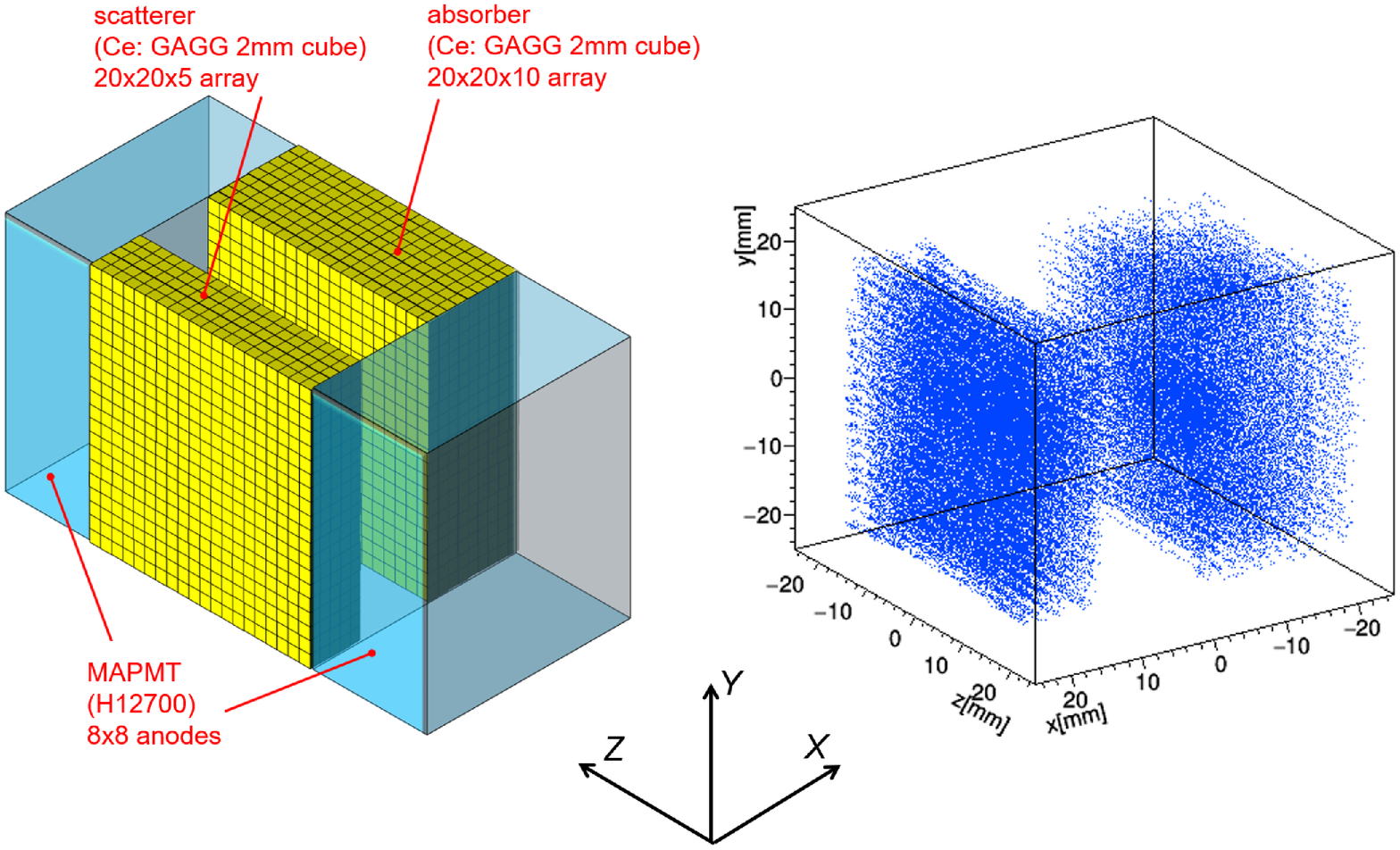}
\caption{
($left$) A schematic configuration of the 3D-PSCC developed in this paper. Pixels in 2~mm cubic Ce:GAGG scintillators accumulate in the 3-D array for both the scatterer and absorber. ($right$) The position map obtained by illuminating a $^{60}$Co source as calculated with Eq. (1).
}
\label{fig:detector}
\end{figure}

\section*{Discussion}
\subsection*{Nuclear cross sections}
Although various nuclear emission lines were observed in the prompt gamma ray spectra shown in Fig.1 and Table 1, it is still controversial which line or energy band is most suitable for online proton therapy monitoring. Ideally, the spatial distributions of prompt gamma rays should resemble with the proton dose distribution, but the physical processes that produce gamma rays via nuclear reactions are completely different from the energy loss process of protons through the electromagnetic interaction. Obviously, each nuclear reaction has different energy threshold and energy dependence of the cross section. Fig.3 ($left$) compares the nuclear cross sections corresponding to various emission lines, namely the 511~keV, 718~keV, 1022~keV and 4438~keV lines as a function of energy as estimated with the PHITS simulation. As noted above, the 4.4~MeV gamma ray line consists of 4438~keV from $^{12}$C$^{*}$ and 4444~keV from $^{11}$B$^{*}$, but the former cross section is two orders of magnitude larger, hence only that of 4438~keV is plotted in the figure. Also note that the 511~keV gamma rays can be emitted from various positron emitters, such as $^{15}$O, $^{13}$N, and $^{11}$C, but emission from $^{15}$O is dominant during proton irradiation\cite{Masuda:2018dg}. Clearly, the energy threshold for a major nuclear reaction that 
emit 4.4~MeV lines, $^{12}$C(p,p)$^{12}$C$^{*}$, is much lower than other nuclear reaction channels; approximately 6~MeV as compared to 14~MeV for $^{16}$O(p,x)$^{15}$O that emit 511~keV, and 25~MeV for $^{12}$C(p,x)$^{10}$B$^{*}$ that emit 718~keV gamma rays. Moreover, its cross section sharply peaks at around 10$-$15~MeV. Note that, the range for such low energy protons is only 1.2$-$2.6~mm in water equivalent pathlength (WEL). This suggests that the 4.4~MeV gamma rays are most effectively emitted just before protons completely stop in the target, and thus may constitute a narrow peak at a position that is very close to the Bragg peak.

\begin{figure}[ht]
\centering
\includegraphics[width=0.9\linewidth]{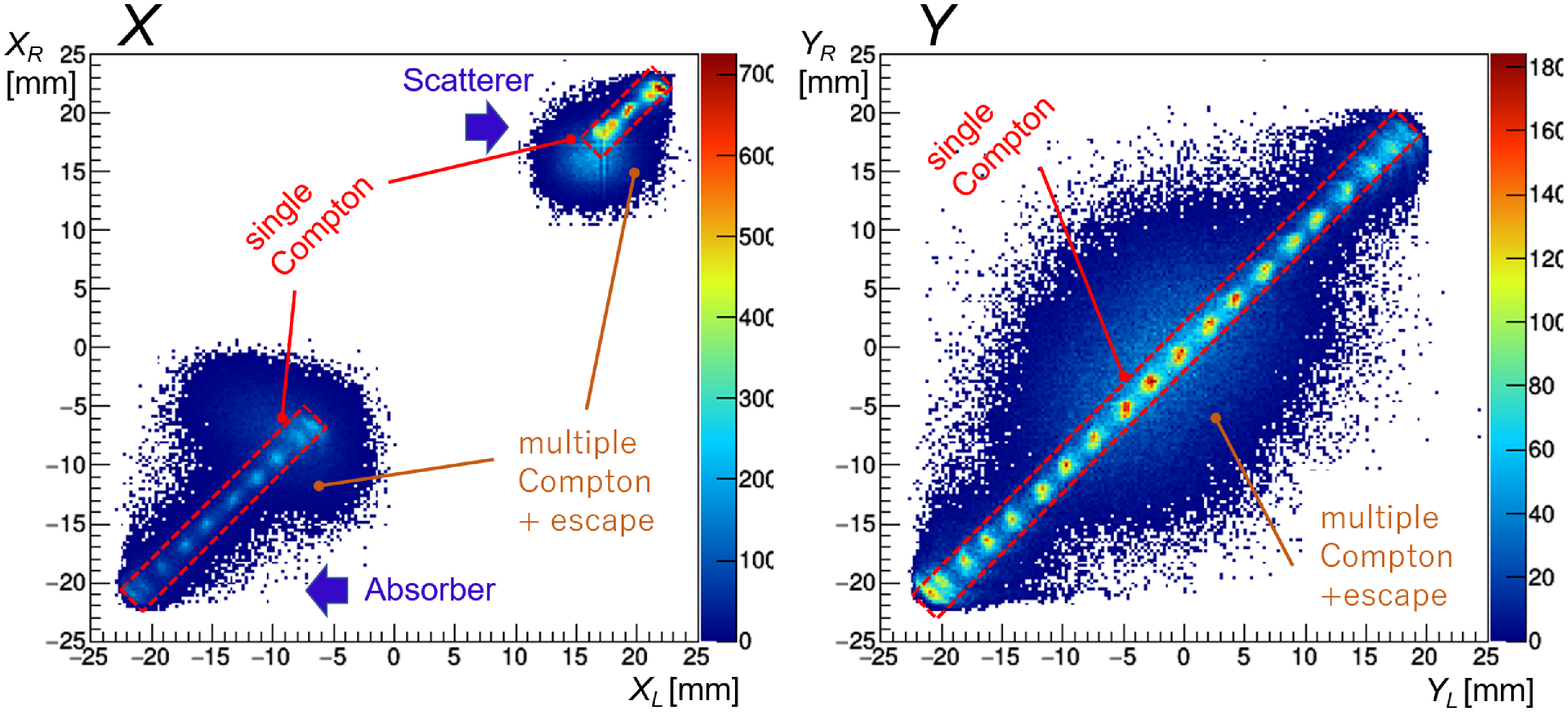}
%
\caption{
($left$) 2-D plot comparing $X_L$ and $X_R$, the $X$-positions calculated from left and right  MAPMTs following the Eqn.(2). Two clumps seen in the upper right and lower left correspond to the gamma ray events in the scatterer and absorber, respectively. The events in the narrow red dashed boxes satisfy $X_L$ $\simeq$ $X_R$, thus corresponding to single Compton events, whereas multiple Compton and/or escape events from pair creation dominate the remaining diffuse regions. ($right$) 2-D plot comparing $Y_L$ and $Y_R$, which again clearly discriminate single and multiple Compton and/or escape events.
}
\label{fig:multiple}
\end{figure}

\subsection*{Anticipated distribution of prompt gamma rays}
Fig.3 ($right$) shows results from a PHITS (ver.2.93\cite{Sato:2018dg}), the 1-D profiles of 511~keV, 718~keV, 1022~keV, and 4.4~MeV lines are obtained by projecting the spatial distributions along the beam axis and are then compared with the proton dose distribution. These simulations clearly support above idea that 4.4~MeV gamma rays are most suitable for proton range verification and dose delivery monitoring. In the experiments using a slit collimator, the peak intensity of the 4.4~MeV gamma rays is approximately coincident with the position of the Bragg peak. However, a slight amount of offset, typically 1$-$4~mm, has been observed even for the same $^{12}$C$^{*}$ peak, depending on chemical composition of the phantom\cite{Kelleter:2017dg}. Our group successfully obtained similar 1-D profiles using a Pb-slit collimator\cite{Koide:2018dg}, but the results regarding Compton imaging of 4.4~MeV gamma rays were not conclusive. The reasons were: (1) the range of 70~MeV protons in PMMA (35~mm) is too short to be resolved with a Compton camera; (2) there is a limited number of 4.4~MeV events that can be used for image reconstruction; (3) multiple Compton, escape, and back-scattering events were misidentified with the preliminary detector configuration. All of these issues have been resolved in this paper.

In the meantime, the proton beam intensity was reduced to approximately 3$-$10~pA and the measurement time was 5~hr. 
In this experiment, the intensity was severely limited by the maximum data acquisition rate of the PMT head amplifies, which is a few kHz when all triggered events in a list-mode were accumulated. The coincidence between the scatterer and absorber is only considered in the offline analysis. To overcome such difficulties, we are developing a new data acquisition system taking hardware coincidence so as to improve the rate tolerance by two orders of magnitude. In that case, we expect that images with the same quality will be obtained within a few minutes for a beam intensity of sub-nA to nA, which is close to the clinical beam. Thus, in the next step, we will use a 200~MeV clinical beam for further image confirmation, and we will also try to image 6.2~MeV gamma rays in real time.

\begin{figure}[ht]
\centering
\includegraphics[width=0.9\linewidth]{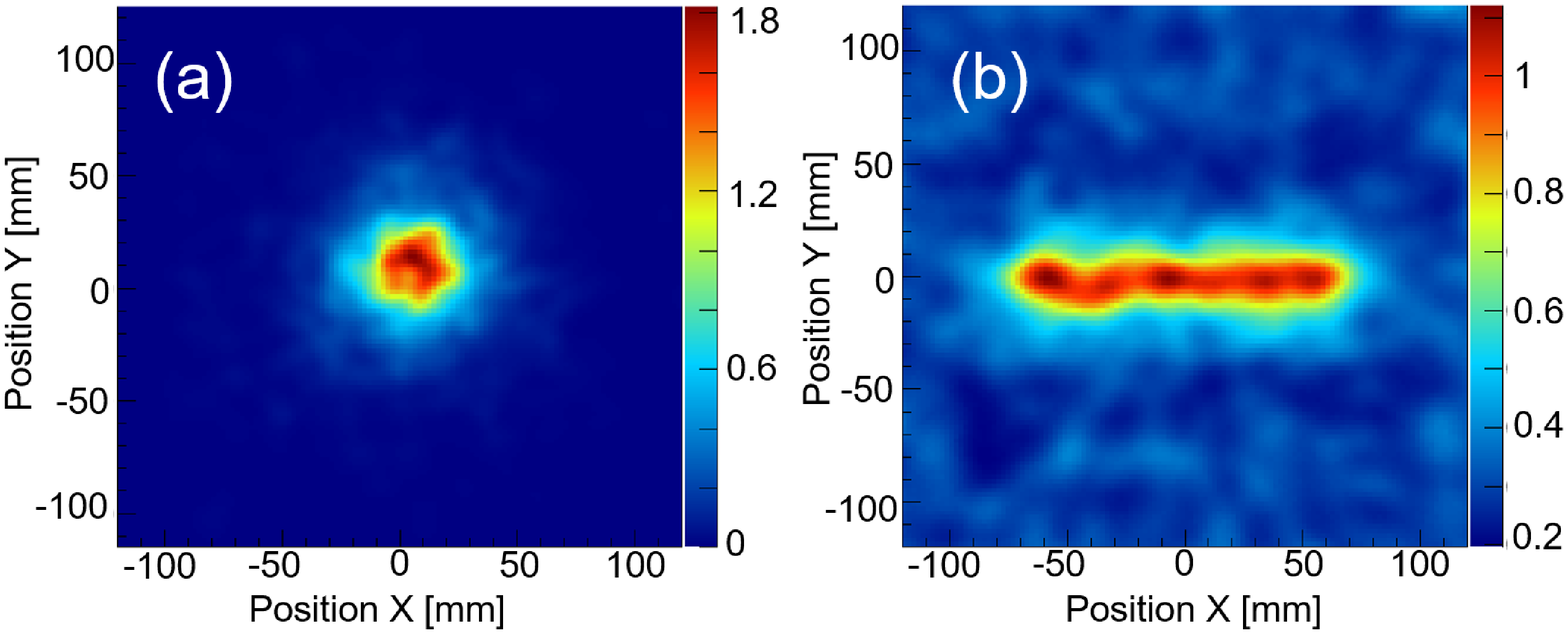}
\caption{
 (a) An experimental image for a 1~MBq $^{60}$Co point source placed 10~cm from the 3D-PSCC. The angular resolution is 9.5$^{\circ}$ (FWHM) at 1.33~MeV.  (b) Simulation image of a 4.4~MeV line gamma ray source whose length is assumed to be 15~cm.
}
\label{fig:image-CC} 
\end{figure}

\section*{Methods}
\subsection*{3D-PSCC for prompt gamma-ray imaging}
We have developed various two-plane Compton cameras consisting of scintillating materials (e.g., Ce;GAGG\cite{Kamada:2012dg}) coupled with multi-pixel photon counters\cite{Kataoka:2013dg}. The great advantage of using a scintillator rather than semiconductor devices like 
Si-CdTe\cite{Takahashi:2004dg,Takeda:2009dg}, CZT\cite{McCleskey:2009dg,Wahl:2015dg,Andrew:2017dg}, or Ge\cite{Motomura:2013dg} detectors is that scintillators are cost effective and have high sensitivity at energies greater than MeV owing to thick and heavy scintillation materials. In the meantime, the scintillator thickness makes the position of the gamma ray interaction uncertain, especially in the depth of interaction (DOI) directions. We therefore implemented 3-D position sensitive scintillators consisting of 2~mm cubic Ce:GAGG scintillators that can measure the gamma ray interaction position within a scintillator in 3-D\cite{Kataoka:2013dg,Kishimoto:2014dg,Kataoka:2015dg}. 
In short, the reflector walls (BaSO$_4$ of 100$\mu$m thickness) divide side-by side crystals in the 2-D ($XY$) direction, while a thin layer of air divides crystals in the DOI ($Z$) direction. Note that conventional dual-sided DOI detectors often use a crystal block consisting of uniform strip scintillators with no divisions in the DOI direction. In contrast, we constructed a crystal block with a polished surface consisting of multiple discrete scintillators to form a substantial gradient of output signals that depend on the DOI positions of the crystals\cite{Kishimoto:2014dg}.

The configuration of the Compton camera is shown in Fig.4 ($left$). The camera consisted of 20$\times$20$\times$5 arrays of Ce:GAGG pixels as a scatterer, whereas 20$\times$20$\times$10 arrays of Ce:GAGG pixels were used as an absorber. 
Unlike other Compton cameras developed in our group\cite{Kataoka:2013dg,Kataoka:2015dg,Kataoka:2018dg,Kishimoto:2017dg}, we used two multi-anode PMTs (MAPMTs; Hamamatsu H12700A) of 8$\times$8 anodes to readout the scintillation light from the right/left ends of the scintillator arrays. This is because the radiation tolerance of the MPPC is uncertain, particularly for damage caused by fast neutrons in proton beam facilities. This matter should be addressed in future work.

\begin{figure}[ht]
\centering
\includegraphics[width=0.9\linewidth]{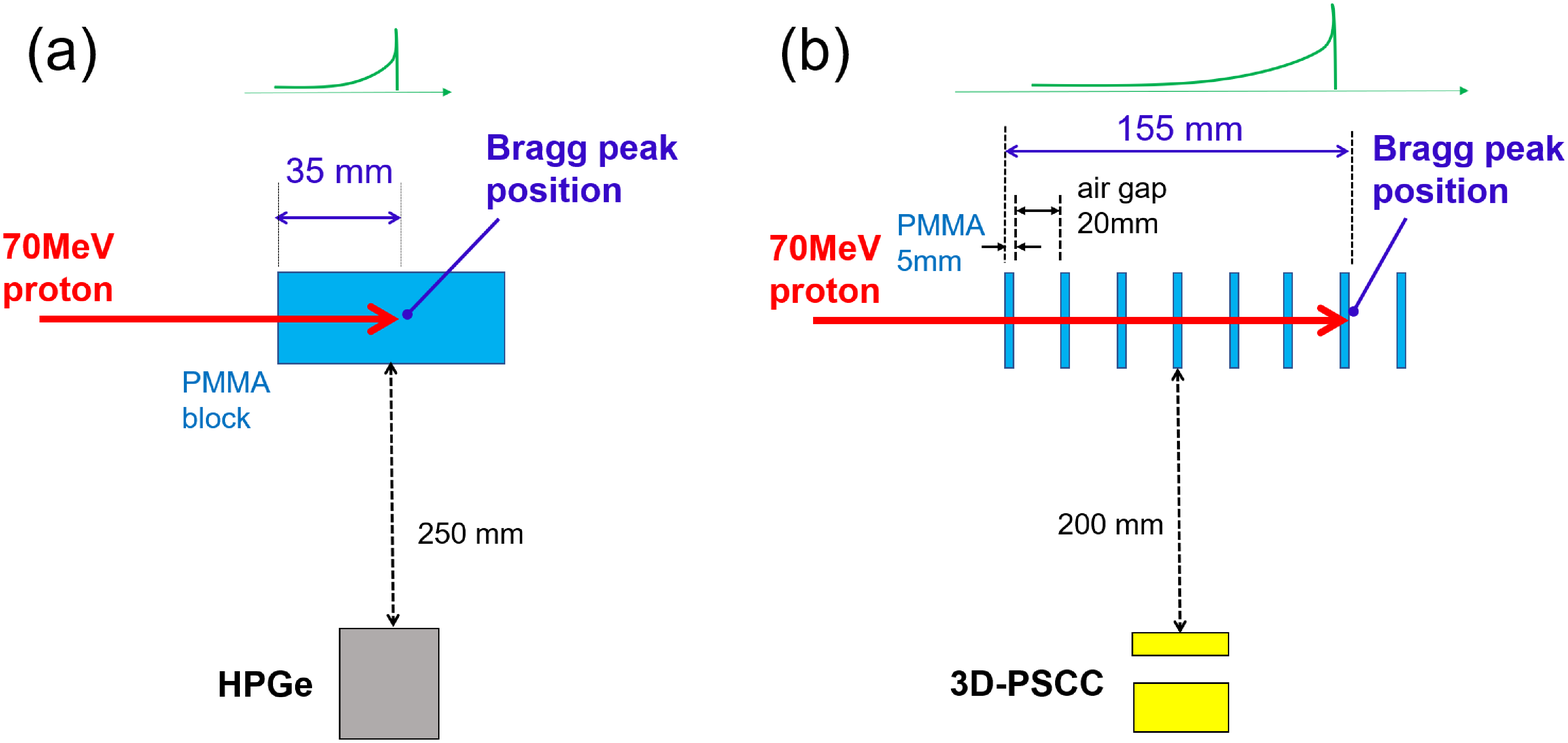}
\caption{
(a) Experimental geometry for measuring prompt gamma rays emitted from a PMMA block phantom using the HPGe detector.  The proton beam intensity was 3 pA throughout the experiment. (b) Experimental geometry for imaging 4.4~MeV gamma rays using the 3D-PSCC. PMMA slab phantoms were irradiated by a 70~MeV proton beam with 3~pA intensity. 
}
\label{fig:setup-HPGE} 
\end{figure}

The output signals from both MAPMTs are fed into the PMT head amplifiers (CLEAR PULSE, 80190) controlled by LabView for position and spectral measurement of each event. An event trigger can be  generated by either head amplifier. The resultant 64$\times$2~ch of pulse-height data (12-bit ADC) is recorded for a time period of 1~$\mu$s following  trigger generation. The output ADC data is compiled, time-stamped in the FPGA and sent to a laptop via a USB~2.0 cable. Note that even a slight difference in gain for each MAPMT and head amplifier channel or in light yield along the DOI direction would affect the spatial resolution of the Compton camera. To correct for this gain, we pre-emptively produced "gain maps" for individual MAPMTs and head amplifiers. We then calibrated the response from all the scintillator pixels using a photo-peak irradiated by a  $^{60}$Co calibration source prior to the Compton imaging. Then the $X$, $Y$, and $Z$ coordinates of each gamma ray interaction position were calculated using the following equations for both the scatterer and absorber:
\begin{equation}
X = \frac{\sum^{N}_{i=1}\{x_i\times(PH_{L,i}+PH_{R,i})\}}{\sum^{N}_{i=1}(PH_{L,i}+PH_{R,i})},\\
Y = \frac{\sum^{N}_{i=1}\{y_i\times(PH_{L,i}+PH_{R,i})\}}{\sum^{N}_{i=1}(PH_{L,i}+PH_{R,i})},\\
Z = \frac{d \times \sum^{N}_{i=1}\ PH_{L,i}}{\sum^{N}_{i=1}(PH_{L,i}+PH_{R,i})},\\
\end{equation}
where $PH_{L,i}$ and $PH_{R,i}$ are the pulse heights, $x_i$ and $y_i$ are the position of $i$-th anode of left and right MAPMTs, $d$ = 42 mm is the DOI length (i.e., length in the $Z$ direction) of the scintillator array. Also, $N$ is the number of MAPMT channels covering either the scatterer or absorber, namely $N$ = 24 for the scatterer and $N$ = 40 for the absorber. The positional map thus calculated for a $^{60}$Co source which emits 1.17 and 1.33~MeV gamma rays is shown in Fig.4 ($right$). 

In this experimental setup, instances of coincidence between the scatterer and the absorber can be easily recorded to detect Compton interaction for the purpose of offline analysis. Since the outputs from all MAPMT channels were recorded in list mode, events during which both the scatterer (24~channels in +6~mm $<$ $X$ $<$ +24~mm) and the absorber (40~channels in $-$24~mm $<$ $X$ $<$ +6~mm) have recorded non-zero ADC values when the coincidence time window was fixed at 1$\mu$s may be selected for the analysis.
During the imaging of high-energy gamma rays, a fraction of multiple Compton and/or escape events within the detector are considered more significant. Based on PHITS simulation estimates, these fractions are $\sim$ 65$\%$ and $\sim$88$\%$ for 
multiple and escape events respectively that occur during the imaging of 4.4~MeV gamma rays. We therefore calculated the $X$ and $Y$ coordinates individually for the left and right MAPMTs as follows for both the scatterer and absorber:
\begin{equation}
X_L = \frac{\sum^{N}_{i=1} (x_i\times PH_{L,i})}{\sum^{N}_{i=1}PH_{L,i}},\\
X_R = \frac{\sum^{N}_{i=1} (x_i\times PH_{R,i})}{\sum^{N}_{i=1}PH_{R,i}},\\
Y_L = \frac{\sum^{N}_{i=1} (y_i\times PH_{L,i})}{\sum^{N}_{i=1}PH_{L,i}},\\
Y_R = \frac{\sum^{N}_{i=1} (y_i\times PH_{R,i})}{\sum^{N}_{i=1}PH_{R,i}}.\\
\end{equation}

Fig.5 shows 2-D histograms comparing $X_L$ vs $X_R$ (Fig.5 $left$) and $Y_L$ vs $Y_R$ (Fig.5 $right$) in the absorber, as an example for a $^{60}$Co source. The events show a narrow correlation, shown as the red-dash boxes in the center. Events outside this area are regarded as multiple Compton and/or escape events. Throughout the paper, we only used single Compton events, in which $X_L$ $\simeq$ $X_R$ and $Y_L$ $\simeq$ $Y_R$ are satisfied in both the scatterer and absorber. Using this simple method for event selection, we confirmed that more than 90$\%$ of multiple Compton events can be eliminated while detecting 4.4~MeV gamma rays. The signal-to-background event ratio would be almost comparable, resulting in a high quality image, as shown in Fig.2.
Also, we applied various energy cuts to remove (1) escape events where part of the incident gamma ray energy is deposited but cannot be totally absorbed, and (2) backscattering events in which gamma rays interact first in the absorber, then backscattered photons are absorbed within the scatterer. Specifically, 65~keV $<$ $E_s$ $<$ 180~keV and 3~MeV $<$ $E_{s}$ + $E_{a}$ $<$ 5~MeV were applied for imaging the 4.4~MeV gamma rays, where $E_{S}$ and $E_{a}$ are the energy deposits in the scatterer and absorber, respectively. 
However, we note that even higher energy gamma rays may deposit a fraction of their energy between 3 and 5~MeV, and then escape from the detector. Such events cannot be rejected in the current flow of event selection, but are estimated to be less than 20$\%$ from the simulation.

\subsection*{Performance verification of 3D-PSCC}
Detailed performance of the 3D-PSCC, such as the angular response and sensitivity to high energy gamma rays, were investigated both in the laboratory experiment and a Geant-4 simulation. The angular resolution of the Compton camera is determined by three factors: (1) position uncertainty, (2) energy uncertainty, and (3) the Doppler broadening effect\cite{Watanabe:2005dg}. In our case, position uncertainty is the primary factor affecting angular resolution and is determined from the side length of the cubic Ce:GAGG scintillator (2~mm throughout the measured energy range). The energy resolution achieved using typical Ce:GAGG pixel in a 3-D scintillator array is 10.5$\%$ (FWHM) when detecting 1.33~MeV gamma rays,  and is anticipated to be 5.7$\%$ for 4.4~MeV gamma rays. The Doppler broadening effect depends on the materials forming the scatterer but has negligibly small influence on angular resolution compared with the other two factors. 
Fig.6 (a) shows the experimental image reconstructed for a $^{60}$Co point source placed 10~cm from the detector. 
We applied the list-mode maximum likelihood expectation maximization (LM-MLEM) algorithm as developed in the literature\cite{Taya:2017dg}, with an iteration number of three. The angular resolution thus measured was 9.5$^{\circ}$ (FWHM) at 1.33~MeV, which is consistent with the simulation. The simulation also predicts the angular resolution improves as energy increases, typically 6.4$^{\circ}$ (FWHM) as measured at 4.4~MeV for prompt gamma ray imaging. 

Also, to correctly reconstruct the geometry and intensity of extended and/or diffuse sources, we first simulated the 3D-PSCC response against a uniformly distributed plane source that covers the entire field of view and emits 4.4~MeV monochromatic gamma rays. The results are used as sensitivity maps to correct the measurement image for extended sources as measured at 4.4~MeV. We also simulated 10~cm, 15~cm, and 20~cm line sources and placed them at 20~cm from the camera to mimic the experimental conditions in the NIRS experiments, as detailed below. From Fig.6 (b), we confirmed that the 3D-PSCC obtained $\pm$10$\%$ uniformity for a 15~cm line source, which emits 4.4~MeV gamma rays. 
 
\subsection*{Setup for measuring prompt gamma-ray spectra}
Fig.7 (a) shows a setup for measuring prompt gamma ray emission in the cyclotron facility at NIRS. We irradiated various phantoms with a 70~MeV proton beam that was narrowly collimated to a diameter of approximately 10~mm. The size of each phantom was sufficiently larger than both the proton range and proton beam diameter, so that all the protons are absorbed in the phantom. For example, we used a PMMA cube phantom with 30$\times$30 mm$^2$ cross section and 10~cm depth along the beam direction, based on a 35~mm proton range for 70~MeV protons in PMMA. The spectrum of prompt gamma rays was gathered with an HPGe detector provided by EG$\&$G (GEM-40190-P-S) by setting the energy range from 10~keV to 5~MeV. The HPGe detector was set perpendicular to the beam axis, and the center of the detector's FOV is approximately aligned to the position of the Bragg peak. The proton beam intensity was reduced to approximately 3~pA so that the probability of a pile up event was less than 5$\%$ during on-beam data acquisition. The output signal from the HPGe was fed into a shaping amplifier (ORTEC 570) with a shaping time set to 6~$\mu$s. Each spectrum was accumulated for 10~min for both the on-beam and off-beam conditions. The distance between the phantom and detector was always fixed to 25~cm.  As shown in Fig.1 ($blue$ spectra), 
a significant amount of gamma-ray emission exists even when the beam is turned off, particularly below 1~MeV. These gamma rays might be  accounted for by the long-lived positron emitters generated in the phantom\cite{Masuda:2018dg} and/or the room background from sources 
such as from activated materials in the cyclotron beam port. This  background is, however, about an order of magnitude smaller than  
the gamma-ray emission we are referring to. For the 3$-$5~MeV gamma rays in particular, contamination is negligibly small -- as can be clearly seen in Fig.1.

\subsection*{Setup for imaging 4.4~MeV gamma rays}
In an actual clinical situation, a proton beam typically has energies ranging from 70~MeV to 250~MeV. Thus, the experimental conditions at the NIRS cyclotron facility using a 70~MeV beam are rather close to the lower limit. However, nuclear reactions that take places in the patient's body and the subsequent production of prompt gamma rays are completely the same, as indicated from the energy threshold of each nuclear reaction and shown in Fig.3 ($left$). However, an obvious difference exists in the proton range. The typical range of clinical proton beam is approximately 25~cm in both water and PMMA, whereas the range is 3.5~cm for 70~MeV protons. Therefore, if we place the 3D-PSCC 20~cm from the proton beam axis, it would be difficult to quantitatively compare with the proton dose distribution, even with the obtained spatial distribution of 4.4~MeV image\cite{Koide:2018dg}. Therefore, we used eight 5~mm thick PMMA slab phantoms that were placed 20~mm apart. A schematic of experimental setup is shown in Fig.7 (b). In this configuration, the proton range extended to 15.5~cm, and the most precise imaging of 4.4~MeV gamma rays can be obtained with such a simple setup. Again, the proton beam intensity was reduced to approximately 3~pA and the measurement time was 5~hr.

\bibliography{sample}

\begin{thebibliography}{10}
\expandafter\ifx\csname url\endcsname\relax
  \def\url#1{\texttt{#1}}\fi
\expandafter\ifx\csname urlprefix\endcsname\relax\def\urlprefix{URL }\fi
\expandafter\ifx\csname doiprefix\endcsname\relax\def\doiprefix{DOI }\fi
\providecommand{\bibinfo}[2]{#2}
\providecommand{\eprint}[2][]{\url{#2}}

\bibitem{Khalil:2011dg}
\bibinfo{author}{Khalil, M.~K.}, \bibinfo{author}{Tremoleda, J.~L.},
  \bibinfo{author}{Bayomy, T.~B.} \& \bibinfo{author}{Gsell, W.}
\newblock \bibinfo{journal}{\bibinfo{title}{{M}olecular {SPECT} {I}maging: {A}n
  {O}verview}}.
\newblock {\emph{\JournalTitle{Int. Jnl. of Molec. Img.}}}
  \textbf{\bibinfo{volume}{2011}}, \bibinfo{pages}{796025}
  (\bibinfo{year}{2011}).

\bibitem{Rosenthal:1995dg}
\bibinfo{author}{Rosenthal, M.~S.} \emph{et~al.}
\newblock \bibinfo{journal}{\bibinfo{title}{Quantitative {SPECT} imaging: a
  review and recommendations by the {F}ocus {C}ommittee of the {S}ociety of
  {N}uclear {M}edicine {C}omputer and {I}nstrumentation {C}ouncil.}}
\newblock {\emph{\JournalTitle{J Nucl Med}}} \textbf{\bibinfo{volume}{36}},
  \bibinfo{pages}{1489--1513} (\bibinfo{year}{1995}).

\bibitem{Kubota:2005dg}
\bibinfo{author}{Kubota, K.}
\newblock \bibinfo{journal}{\bibinfo{title}{From tumor biology to clinical
  {P}et: a review of positron emission tomography ({PET}) in oncology}}.
\newblock {\emph{\JournalTitle{Ann. Nucl. Med}}} \textbf{\bibinfo{volume}{15}},
  \bibinfo{pages}{471--486} (\bibinfo{year}{2005}).

\bibitem{Coleman:2005dg}
\bibinfo{author}{Coleman, R.}
\newblock \bibinfo{journal}{\bibinfo{title}{Positron emission tomography
  diagnosis of {A}lzheimer's disease}}.
\newblock {\emph{\JournalTitle{Neuroimaging Clin N Am}}}
  \textbf{\bibinfo{volume}{15}}, \bibinfo{pages}{837--846}
  (\bibinfo{year}{2005}).

\bibitem{Weilhammer:2000dg}
\bibinfo{author}{Weilhammer, P.}
\newblock \bibinfo{journal}{\bibinfo{title}{Overview: silicon vertex detectors
  and trackers}}.
\newblock {\emph{\JournalTitle{Nucl. Instrm. Methods A}}}
  \textbf{\bibinfo{volume}{453}}, \bibinfo{pages}{60--70}
  (\bibinfo{year}{2000}).

\bibitem{Torii:2015dg}
\bibinfo{author}{Torii, S.} \emph{et~al.}
\newblock \bibinfo{journal}{\bibinfo{title}{The {CAL}orimetric {E}lectron
  {T}elescope ({CALET}): {H}igh {E}nergy {A}stroparticle {P}hysics
  {O}bservatory on the {I}nternational {S}pace {S}tation}}.
\newblock {\emph{\JournalTitle{Proceedings of Science, ICRC2015}}}
  \textbf{\bibinfo{volume}{34}}, \bibinfo{pages}{id.581}
  (\bibinfo{year}{2015}).

\bibitem{Knodlseder:2016dg}
\bibinfo{author}{J\"{u}rgen, K.}
\newblock \bibinfo{journal}{\bibinfo{title}{The future of gamma-ray
  astronomy}}.
\newblock {\emph{\JournalTitle{C. R. Physique}}} \textbf{\bibinfo{volume}{17}},
  \bibinfo{pages}{663--678} (\bibinfo{year}{2016}).

\bibitem{Atwood:2009dg}
\bibinfo{author}{Atwood, W.~B.} \emph{et~al.}
\newblock \bibinfo{journal}{\bibinfo{title}{The {L}arge {A}rea {T}elescope on
  the {F}ermi {G}amma-{R}ay {S}pace {T}elescope {M}ission}}.
\newblock {\emph{\JournalTitle{Astrophys. J.}}} \textbf{\bibinfo{volume}{697}},
  \bibinfo{pages}{1071--1102} (\bibinfo{year}{2009}).

\bibitem{Acero:2015dg}
\bibinfo{author}{Acero, F.} \emph{et~al.}
\newblock \bibinfo{journal}{\bibinfo{title}{{F}ermi {L}arge {A}rea {T}elescope
  {T}hird {S}ource {C}atalog}}.
\newblock {\emph{\JournalTitle{Astrophys. J. Suppl. Se.}}}
  \textbf{\bibinfo{volume}{218}}, \bibinfo{pages}{23--64}
  (\bibinfo{year}{2015}).

\bibitem{shon:1973dg}
\bibinfo{author}{Sch\"{o}nfelder, V.}, \bibinfo{author}{Hirner, A.} \&
  \bibinfo{author}{Schneider, K.}
\newblock \bibinfo{journal}{\bibinfo{title}{A telescope for soft gamma ray
  astronomy}}.
\newblock {\emph{\JournalTitle{Nucl. Instrm. Methods A}}}
  \textbf{\bibinfo{volume}{107}}, \bibinfo{pages}{385--394}
  (\bibinfo{year}{1973}).

\bibitem{Lopes:2015dg}
\bibinfo{author}{Lopes, P.~C.} \emph{et~al.}
\newblock \bibinfo{journal}{\bibinfo{title}{Time-resolved imaging of
  prompt-gamma rays for proton range verification using a knife-edge slit
  camera based on digital photon counters}}.
\newblock {\emph{\JournalTitle{Phys. Med. Biol.}}}
  \textbf{\bibinfo{volume}{60}}, \bibinfo{pages}{6063--6085}
  (\bibinfo{year}{2015}).

\bibitem{Golnik:2016dg}
\bibinfo{author}{Golnik, C.} \emph{et~al.}
\newblock \bibinfo{journal}{\bibinfo{title}{Tests of a {C}ompton imaging
  prototype in a monoenergetic 4.44~{M}ev photon filed - a benchmark setup for
  prompt gamma-ray imaging devices}}.
\newblock {\emph{\JournalTitle{Journal of Instrumentation}}}
  \textbf{\bibinfo{volume}{11}}, \bibinfo{pages}{P06009}
  (\bibinfo{year}{2016}).

\bibitem{Hilaire:2016dg}
\bibinfo{author}{Hilaire, E.}, \bibinfo{author}{Sarrut, D.},
  \bibinfo{author}{Peyrin, F.} \& \bibinfo{author}{Maxim, V.}
\newblock \bibinfo{journal}{\bibinfo{title}{Proton therapy monitoring by
  {C}ompton imaging: influence of the large energy spectrum of the
  prompt-$\gamma$ radiation}}.
\newblock {\emph{\JournalTitle{Phys. Med. Biol.}}}
  \textbf{\bibinfo{volume}{61}}, \bibinfo{pages}{3--368}
  (\bibinfo{year}{2016}).

\bibitem{Kelleter:2017dg}
\bibinfo{author}{Kelleter, L.} \emph{et~al.}
\newblock \bibinfo{journal}{\bibinfo{title}{Spectroscopic study of prompt-gamma
  emission for range verification in proton therapy}}.
\newblock {\emph{\JournalTitle{Phys. Med.}}} \textbf{\bibinfo{volume}{34}},
  \bibinfo{pages}{7--17} (\bibinfo{year}{2017}).

\bibitem{Koide:2018dg}
\bibinfo{author}{Koide, A.} \emph{et~al.}
\newblock \bibinfo{journal}{\bibinfo{title}{Spatially resolved measurement of
  wideband prompt gamma-ray emission toward on-line monitor for the future
  proton therapy}}.
\newblock {\emph{\JournalTitle{Nucl. Instrm. Methods A}}}
  \textbf{\bibinfo{volume}{in press}} (\bibinfo{year}{2018}).

\bibitem{Min:2006dg}
\bibinfo{author}{Min, C.-H.}, \bibinfo{author}{Kim, C.~H.},
  \bibinfo{author}{Youn, M.-Y.} \& \bibinfo{author}{Kim, J.-W.}
\newblock \bibinfo{journal}{\bibinfo{title}{Prompt gamma measurements for
  locating the dose falloff region in the proton therapy}}.
\newblock {\emph{\JournalTitle{Applied Physics Letters}}}
  \textbf{\bibinfo{volume}{89}}, \bibinfo{pages}{183517}
  (\bibinfo{year}{2006}).

\bibitem{Knopf:2013dg}
\bibinfo{author}{Knopf, A.-C.} \& \bibinfo{author}{Romax, A.}
\newblock \bibinfo{journal}{\bibinfo{title}{In vivo proton range verification:
  a review}}.
\newblock {\emph{\JournalTitle{Phys. Med. Biol.}}}
  \textbf{\bibinfo{volume}{58}}, \bibinfo{pages}{R131--R160}
  (\bibinfo{year}{2013}).

\bibitem{Kurosawa:2012dg}
\bibinfo{author}{Kurosawa, S.} \emph{et~al.}
\newblock \bibinfo{journal}{\bibinfo{title}{Prompt gamma detection for range
  verification in proton therapy}}.
\newblock {\emph{\JournalTitle{Current Applied Physics}}}
  \textbf{\bibinfo{volume}{12}}, \bibinfo{pages}{364--368}
  (\bibinfo{year}{2012}).

\bibitem{Nishio:2010dg}
\bibinfo{author}{Nishio, T.} \emph{et~al.}
\newblock \bibinfo{journal}{\bibinfo{title}{The {D}evelopment and {C}linical
  {U}se of a {B}eam {ON}-{LINE} {PET} {S}ystem {M}ounted on a {R}otating
  {G}antry {P}ort in {P}roton {T}herapy}}.
\newblock {\emph{\JournalTitle{Radiation Oncology}}}
  \textbf{\bibinfo{volume}{76}}, \bibinfo{pages}{277--286}
  (\bibinfo{year}{2010}).

\bibitem{Taya:2016dg}
\bibinfo{author}{Taya, T.} \emph{et~al.}
\newblock \bibinfo{journal}{\bibinfo{title}{First demonstration of real-time
  gamma imaging by using a handheld {C}ompton camera for particle therapy}}.
\newblock {\emph{\JournalTitle{Nucl. Instrm. Methods A}}}
  \textbf{\bibinfo{volume}{831}}, \bibinfo{pages}{355--361}
  (\bibinfo{year}{2016}).

\bibitem{Taya:2017dg}
\bibinfo{author}{Taya, T.} \emph{et~al.}
\newblock \bibinfo{journal}{\bibinfo{title}{Optimization and verification of
  image reconstruction for a {C}ompton camera towards application as an on-line
  monitor for particle therapy}}.
\newblock {\emph{\JournalTitle{Journal of Instrumentation}}}
  \textbf{\bibinfo{volume}{12}}, \bibinfo{pages}{P07015}
  (\bibinfo{year}{2017}).

\bibitem{Kozol:2002dg}
\bibinfo{author}{Kozolovsky, B.}, \bibinfo{author}{Murphy, R.~J.} \&
  \bibinfo{author}{Ramaty, R.}
\newblock \bibinfo{journal}{\bibinfo{title}{Nuclear deexcitation gamma-ray
  lines from accelerated particle interactions}}.
\newblock {\emph{\JournalTitle{Astrophys. J. Suppl. S.}}}
  \textbf{\bibinfo{volume}{141}}, \bibinfo{pages}{523--541}
  (\bibinfo{year}{2002}).

\bibitem{Kobayashi:2010dg}
\bibinfo{author}{Kobayashi, S.} \emph{et~al.}
\newblock \bibinfo{journal}{\bibinfo{title}{Determining the {A}bsolute
  {A}bundances of {N}atural {R}adioactive {E}lements on the {L}unar {S}urface
  by the {K}aguya {G}amma-ray {S}pectrometer}}.
\newblock {\emph{\JournalTitle{Space Science Reviews}}}
  \textbf{\bibinfo{volume}{154}}, \bibinfo{pages}{193--218}
  (\bibinfo{year}{2002}).

\bibitem{Vestrand:1999dg}
\bibinfo{author}{Vestrand, W.~T.} \emph{et~al.}
\newblock \bibinfo{journal}{\bibinfo{title}{{THE} {SOLAR} {MAXIMUM} {MISSION}
  {ATLAS} {OF} {GAMMA}-{RAY} {FLARES}}}.
\newblock {\emph{\JournalTitle{Astrophys. J. Suppl. S.}}}
  \textbf{\bibinfo{volume}{120}}, \bibinfo{pages}{409--467}
  (\bibinfo{year}{1999}).

\bibitem{Iyudin:1994dg}
\bibinfo{author}{Iyudin, A.~F.} \emph{et~al.}
\newblock \bibinfo{journal}{\bibinfo{title}{{COMPTEL} observations of {Ti}-44
  gamma-ray line emission from {CAS A}}}.
\newblock {\emph{\JournalTitle{Astron. $\&$ Astrophys.}}}
  \textbf{\bibinfo{volume}{284}}, \bibinfo{pages}{L1--L4}
  (\bibinfo{year}{1994}).

\bibitem{Diehl:1995dg}
\bibinfo{author}{Diehl, R.} \emph{et~al.}
\newblock \bibinfo{journal}{\bibinfo{title}{{COMPTEL} observations of
  {G}alactic $^{26}${Al} emission}}.
\newblock {\emph{\JournalTitle{Astron. $\&$ Astrophys.}}}
  \textbf{\bibinfo{volume}{298}}, \bibinfo{pages}{445--460}
  (\bibinfo{year}{1995}).

\bibitem{Dogiel:2009dg}
\bibinfo{author}{Dogiel, V.~A.} \emph{et~al.}
\newblock \bibinfo{journal}{\bibinfo{title}{Nuclear interaction gamma-ray lines
  from the {G}alactic center region}}.
\newblock {\emph{\JournalTitle{Astron. $\&$ Astrophys.}}}
  \textbf{\bibinfo{volume}{508}}, \bibinfo{pages}{1--7} (\bibinfo{year}{2009}).

\bibitem{Beekman:2007dg}
\bibinfo{author}{Beekman, F.~J.} \& \bibinfo{author}{der Have, F.~V.}
\newblock \bibinfo{journal}{\bibinfo{title}{The pinhole: gateway to
  ultra-high-resolution three-dimensional radionuclide imaging}}.
\newblock {\emph{\JournalTitle{J. Nucl. Med. Mol. Imaging}}}
  \textbf{\bibinfo{volume}{34}}, \bibinfo{pages}{151--161}
  (\bibinfo{year}{2007}).

\bibitem{Caroli:1987dg}
\bibinfo{author}{Caroli, E.}, \bibinfo{author}{Stephen, J.~B.},
  \bibinfo{author}{Di~Cocco, G.}, \bibinfo{author}{Natalucci, L.} \&
  \bibinfo{author}{Spizzichino, A.}
\newblock \bibinfo{journal}{\bibinfo{title}{Coded aperture imaging in {X}- and
  gamma-ray astronomy}}.
\newblock {\emph{\JournalTitle{Space Science Reviews}}}
  \textbf{\bibinfo{volume}{45}}, \bibinfo{pages}{349--403}
  (\bibinfo{year}{1987}).

\bibitem{shon:1993dg}
\bibinfo{author}{Sch\"{o}nfelder, V.} \emph{et~al.}
\newblock \bibinfo{journal}{\bibinfo{title}{Instrument description and
  performance of the imaging gamma-ray telescope {COMPTEL} aboard the {Compton}
  {G}amma-{R}ay {Observatory}}}.
\newblock {\emph{\JournalTitle{Astrophys. J. Suppl. S.}}}
  \textbf{\bibinfo{volume}{86}}, \bibinfo{pages}{675--692}
  (\bibinfo{year}{1993}).

\bibitem{Tanimori:2004dg}
\bibinfo{author}{Tanimori, T.} \emph{et~al.}
\newblock \bibinfo{journal}{\bibinfo{title}{Mev $\gamma$-ray imaging detector
  with micro-{TPC}}}.
\newblock {\emph{\JournalTitle{New Astronomy Reviews}}}
  \textbf{\bibinfo{volume}{48}}, \bibinfo{pages}{263--268}
  (\bibinfo{year}{2004}).

\bibitem{Takahashi:2004dg}
\bibinfo{author}{Takahashi, T.} \emph{et~al.}
\newblock \bibinfo{journal}{\bibinfo{title}{Hard {X}-ray and $\gamma$-ray
  detectors for the {N}e{XT} mission}}.
\newblock {\emph{\JournalTitle{New Astronomy Reviews}}}
  \textbf{\bibinfo{volume}{48}}, \bibinfo{pages}{269--273}
  (\bibinfo{year}{2004}).

\bibitem{Kataoka:2013dg}
\bibinfo{author}{Kataoka, J.} \emph{et~al.}
\newblock \bibinfo{journal}{\bibinfo{title}{Handy {C}ompton camera using 3{D}
  position-sensitive scintillators coupled with large-area monolithic {MPPC}
  arrays}}.
\newblock {\emph{\JournalTitle{Nucl. Instrm. Methods A}}}
  \textbf{\bibinfo{volume}{541}}, \bibinfo{pages}{398--404}
  (\bibinfo{year}{2013}).

\bibitem{Motomura:2013dg}
\bibinfo{author}{Motomura, S.} \emph{et~al.}
\newblock \bibinfo{journal}{\bibinfo{title}{Improved imaging performance of a
  semiconductor {C}ompton camera {GREI} makes for a new methodology to
  integrate bio-metal analysis and molecular imaging technology in living
  organisms}}.
\newblock {\emph{\JournalTitle{Jnl. of Anal. Atom. Spectrom.}}}
  \textbf{\bibinfo{volume}{28}}, \bibinfo{pages}{934--939}
  (\bibinfo{year}{2017}).

\bibitem{Tanimori:2017dg}
\bibinfo{author}{Tanimori, T.} \emph{et~al.}
\newblock \bibinfo{journal}{\bibinfo{title}{Establishment of {I}maging
  {S}pectroscopy of {N}uclear {G}amma-{R}ays based on {G}eometrical {O}ptics}}.
\newblock {\emph{\JournalTitle{Scientific Reports}}}
  \textbf{\bibinfo{volume}{7}}, \bibinfo{pages}{41511} (\bibinfo{year}{2017}).

\bibitem{Tod:1974dg}
\bibinfo{author}{Todd, R.~W.}, \bibinfo{author}{Nightingale, J.~M.} \&
  \bibinfo{author}{Everett, D.~B.}
\newblock \bibinfo{journal}{\bibinfo{title}{A proposed $\gamma$ camera}}.
\newblock {\emph{\JournalTitle{Nature}}} \textbf{\bibinfo{volume}{251}},
  \bibinfo{pages}{132--134} (\bibinfo{year}{1974}).

\bibitem{Kishimoto:2014dg}
\bibinfo{author}{Kishimoto, A.} \emph{et~al.}
\newblock \bibinfo{journal}{\bibinfo{title}{First demonstration of multi-color
  3-{D} in vivo imaging using ultra-compact {C}ompton camera}}.
\newblock {\emph{\JournalTitle{Journal of Instrumentation}}}
  \textbf{\bibinfo{volume}{9}}, \bibinfo{pages}{P11025} (\bibinfo{year}{2014}).

\bibitem{Wahl:2015dg}
\bibinfo{author}{Wahl, C.~G.} \emph{et~al.}
\newblock \bibinfo{journal}{\bibinfo{title}{The {P}olaris-{H} imaging
  spectrometer}}.
\newblock {\emph{\JournalTitle{Nucl. Instrm. Methods A}}}
  \textbf{\bibinfo{volume}{784}}, \bibinfo{pages}{377--381}
  (\bibinfo{year}{2015}).

\bibitem{Takeda:2015dg}
\bibinfo{author}{Takeda, S.} \emph{et~al.}
\newblock \bibinfo{journal}{\bibinfo{title}{A portable {S}i/{C}d{T}e {C}ompton
  camera and its applications to the visualization of radioactive substances}}.
\newblock {\emph{\JournalTitle{Nucl. Instrm. Methods A}}}
  \textbf{\bibinfo{volume}{787}}, \bibinfo{pages}{207--211}
  (\bibinfo{year}{2015}).

\bibitem{Tomono:2017dg}
\bibinfo{author}{Tomono, D.} \emph{et~al.}
\newblock \bibinfo{journal}{\bibinfo{title}{First {O}n-site {T}rue
  {G}amma-{R}ay {I}maging-{S}pectroscopy of {C}ontamination near {F}ukushima
  {P}lant}}.
\newblock {\emph{\JournalTitle{Scientific Reports}}}
  \textbf{\bibinfo{volume}{7}}, \bibinfo{pages}{41972} (\bibinfo{year}{2017}).

\bibitem{Mochizuki:2017dg}
\bibinfo{author}{Mochizuki, S.} \emph{et~al.}
\newblock \bibinfo{journal}{\bibinfo{title}{First demonstration of aerial
  gamma-ray imaging using drone for prompt radiation survey in {F}ukushima}}.
\newblock {\emph{\JournalTitle{Journal of Instrumentation}}}
  \textbf{\bibinfo{volume}{12}}, \bibinfo{pages}{P11014}
  (\bibinfo{year}{2017}).

\bibitem{Andrew:2017dg}
\bibinfo{author}{Haefner, A.} \emph{et~al.}
\newblock \bibinfo{journal}{\bibinfo{title}{Handheld real-time volumetric 3-{D}
  gamma-ray imaging}}.
\newblock {\emph{\JournalTitle{Nucl. Instrm. Methods A}}}
  \textbf{\bibinfo{volume}{857}}, \bibinfo{pages}{42--49}
  (\bibinfo{year}{2017}).

\bibitem{Sato:2018dg}
\bibinfo{author}{Sato, T.} \emph{et~al.}
\newblock \bibinfo{journal}{\bibinfo{title}{Features of {P}article and {H}eavy
  {I}on {T}ransport code system ({PHITS}) version 3.02}}.
\newblock {\emph{\JournalTitle{J. Nucl. Sci. Technol.}}}
  \bibinfo{pages}{https://doi.org/10.1080/00223131.2017.1419890}
  (\bibinfo{year}{2018}).

\bibitem{Masuda:2018dg}
\bibinfo{author}{Masuda, T.} \emph{et~al.}
\newblock \bibinfo{journal}{\bibinfo{title}{Measurement of nuclear reaction
  cross sections by using {C}herenkov radiation toward high-precision proton
  therapy}}.
\newblock {\emph{\JournalTitle{Scientific Reports}}}
  \textbf{\bibinfo{volume}{8}}, \bibinfo{pages}{2570} (\bibinfo{year}{2018}).

\bibitem{Kamada:2012dg}
\bibinfo{author}{Kamada, K.} \emph{et~al.}
\newblock \bibinfo{journal}{\bibinfo{title}{2 inch diameter single crystal
  growth and scintillation properties of
  {C}e:{G}d$_3${A}l$_2${G}a$_3${O}$_{12}$}}.
\newblock {\emph{\JournalTitle{Journal of Crystal Growth}}}
  \textbf{\bibinfo{volume}{352}}, \bibinfo{pages}{88--90}
  (\bibinfo{year}{2012}).

\bibitem{Takeda:2009dg}
\bibinfo{author}{Takeda, S.} \emph{et~al.}
\newblock \bibinfo{journal}{\bibinfo{title}{Experimental {R}esults of the
  {G}amma-{R}ay {I}maging {C}apability with a {S}i/{C}d{T}e {S}emiconductor
  {C}ompton {C}amera}}.
\newblock {\emph{\JournalTitle{IEEE Trans. on Nucl. Sci.}}}
  \textbf{\bibinfo{volume}{56}}, \bibinfo{pages}{783--790}
  (\bibinfo{year}{2009}).

\bibitem{McCleskey:2009dg}
\bibinfo{author}{McCleskey, M.} \emph{et~al.}
\newblock \bibinfo{journal}{\bibinfo{title}{Evaluation of a multistage
  {C}d{Z}n{T}e compton camera for prompt $\gamma$ imaging for proton therapy}}.
\newblock {\emph{\JournalTitle{Nucl. Instrm. Methods A}}}
  \textbf{\bibinfo{volume}{785}}, \bibinfo{pages}{163--169}
  (\bibinfo{year}{2015}).

\bibitem{Kataoka:2015dg}
\bibinfo{author}{Kataoka, J.} \emph{et~al.}
\newblock \bibinfo{journal}{\bibinfo{title}{Recent progress of {MPPC}-based
  scintillation detectors in high precision {X}-ray and gamma-ray imaging}}.
\newblock {\emph{\JournalTitle{Nucl. Instrm. Methods A}}}
  \textbf{\bibinfo{volume}{784}}, \bibinfo{pages}{248--254}
  (\bibinfo{year}{2015}).

\bibitem{Kataoka:2018dg}
\bibinfo{author}{Kataoka, J.} \emph{et~al.}
\newblock \bibinfo{journal}{\bibinfo{title}{Ultracompact {C}ompton camera for
  innovative gamma-ray imaging}}.
\newblock {\emph{\JournalTitle{Nucl. Instrm. Methods A}}}
  \textbf{\bibinfo{volume}{in press}} (\bibinfo{year}{2018}).

\bibitem{Kishimoto:2017dg}
\bibinfo{author}{Kishimoto, A.} \emph{et~al.}
\newblock \bibinfo{journal}{\bibinfo{title}{First demonstration of multi-color
  3-{D} in vivo imaging using ultra-compact {C}ompton camera}}.
\newblock {\emph{\JournalTitle{Scientific Reports}}}
  \textbf{\bibinfo{volume}{7}}, \bibinfo{pages}{2110} (\bibinfo{year}{2017}).

\bibitem{Watanabe:2005dg}
\bibinfo{author}{Watanabe, S.} \emph{et~al.}
\newblock \bibinfo{journal}{\bibinfo{title}{A {S}i/{C}dte semiconductor
  {C}ompton camera}}.
\newblock {\emph{\JournalTitle{IEEE Trans. on Nucl. Sci.}}}
  \textbf{\bibinfo{volume}{52}}, \bibinfo{pages}{2045--2051}
  (\bibinfo{year}{2005}).

\end{thebibliography}

\section*{Acknowledgements}

This research was supported by JSPS KAKENHI Grant Number
JP15H05720. Careful supports by technical staffs in the cyclotron facility in  NIRS is also acknowledged to complete the experiments presented in the paper.

\section*{Author contributions statement}
J.K. conceived the concept of this research. A.K., J.K., T.T., S.M. K.S. L.T. K.F. T.M. and T.K. conceived the experiments, T.M. derived various cross sections for major nuclear reactions, T.I. provided the technical advices for the NIRS experiments and interpretation of results. J.K. wrote the paper, and ll authors reviewed the manuscript. 

\section*{Additional information}
\textbf{Competing financial interests:} 
The authors declare that they have no competing interests.


\end{document}